\documentclass[preprint2]{aastex}

\newcommand{\Mj}{$M_{\rm Jup}$}
\newcommand{\kms}{km\,s$^{-1}$}
\newcommand{\vsini}{$v_{\rm rot}$\,sin\,$i$}

\slugcomment{Revised version (April 2006)}

\shorttitle{Spectroscopic rotational velocities of brown dwarfs}
\shortauthors{Zapatero Osorio et al.}

\begin{document}

\title{Spectroscopic rotational velocities of brown dwarfs}

\author{M. R. Zapatero Osorio\altaffilmark{1}}
\affil{LAEFF-INTA, P.\,O$.$ 50727, E-28080 Madrid, Spain}
\email{mosorio@laeff.inta.es}

\author{E. L. Mart\'\i n\altaffilmark{2} and H. Bouy}
\affil{IAC, V\'\i a L\'actea s/n, E-38205 La Laguna, Tenerife, Spain}
\email{ege@iac.es, bouy@iac.es}

\author{R. Tata, R. Deshpande}
\affil{University of Central Florida, Dept$.$ of Physics, P.\,O$.$ 
       162385, Orlando FL 32816-2385, USA}
\email{tata@physics.ucf.edu, rohit@physics.ucf.edu}

\and

\author{R. J. Wainscoat}
\affil{Institute for Astronomy, 2680 Woodlawn Drive, Honolulu, HI 96822, USA}
\email{rjw@ifa.hawaii.edu}

\altaffiltext{1}{Also at IAC, V\'\i a L\'actea s/n, E-38205 La Laguna, 
       Tenerife, Spain}
\altaffiltext{2}{Also at University of Central Florida, Dept$.$ of Physics, 
       P.\,O$.$ 162385, Orlando FL 32816-2385, USA}

\begin{abstract}
  We have obtained projected rotation velocities (\vsini) of a sample
  of 19 ultracool dwarfs with spectral types in the interval M6.5--T8
  using high-resolution, near-infrared spectra obtained with NIRSPEC
  and the Keck II telescope. Among our targets there are two young
  brown dwarfs, two likely field stars, and fifteen likely brown
  dwarfs (30--72\Mj) of the solar neighborhood. Our results indicate
  that the T-type dwarfs are fast rotators in marked contrast to
  M-type stars. We have derived \vsini~velocities between $\le$15 and
  40~\kms~ for them, and have found no clear evidence for T dwarfs
  rotating strongly faster than L dwarfs. However, there is a hint for
  an increasing lower envelope on moving from mid-M to the L spectral
  types in the \vsini--spectral type diagram that was previously
  reported in the literature; our \vsini~results extend it to even
  cooler types.  Assuming that field brown dwarfs have a size of
  0.08--0.1~$R_{\odot}$, we can place an upper limit of 12.5 h on the
  equatorial rotation period of T-type brown dwarfs. In addition, we
  have compared our \vsini~measurements to spectroscopic rotational
  velocities of very young brown dwarfs of similar mass available in
  the literature. The comparison, although model-dependent, suggests
  that brown dwarfs lose some angular momentum during their
  contraction; however, their spin down time seems to be significantly
  longer than that of solar-type to early-M stars.

\end{abstract}

\keywords{stars: low-mass, brown dwarfs ---  stars: rotation --- 
individual (\objectname[Pleiades]{PPl 1},
            \objectname{2MASS J03341218-4953322},
            \objectname{vB 10},
            \objectname{LP 944-20},
            \objectname{2MASS J00361617+1821104},
            \objectname{2MASS J22244381-0158521},
            \objectname{SDSS J053951.99-005902.0},
            \objectname{2MASS J17281150+3948593AB},
            \objectname{2MASS J16322911+1904407},
            \objectname{DENIS-P J0255.0-4700},
            \objectname{SDSSp J125453.90-012247.4},
            \objectname{2MASS J05591914-1404488},
            \objectname{2MASS J15031961+2525196},
            \objectname{SDSS J134646.45-003150.4},
            \objectname{SDSS J162414.37+002915.6},
            \objectname{2MASS J15530228+1532369AB},
            \objectname{2MASS J12171110-0311131},
            \objectname{GL 570D},
            \objectname{2MASS J04151954-0935066}
)}

\section{Introduction}
Rotation is a key parameter fundamental in examining stellar activity
and angular momentum evolution. The rotation of stars has been studied
for nearly all spectral types and for a wide range of ages. These
studies have contributed to our knowledge of the stellar angular
momentum history. In particular, field early-M dwarfs are amongst the
slowest rotating stars on the main sequence with typical projected
rotational velocities (\vsini) below 5~\kms~(e.g., Marcy \& Chen
\cite{marcy92}; Delfosse et al. \cite{delfosse98}). It has been
suggested that many of these stars lose about 98\%~of their angular
momentum during their lifetimes via magnetic activity, winds and mass
loss. However, very little is known about the rotation rate and
angular momentum evolution of brown dwarfs.

The state-of-the-art evolutionary models by Baraffe et al.
\cite{baraffe03} and Burrows et al. \cite{burrows97} predict that a
star at the substellar mass limit (0.072\,$M_{\odot}$, or
$\sim$72\,\Mj) reaches the hydrostatic equilibrium at an effective
temperature of about 1800\,K. However, T-type dwarfs (for the spectral
definition see Burgasser et al. \cite{burgasser02a} and Geballe et al.
\cite{geballe02}) are characterized by surface temperatures cooler
than $\sim$1300\,K (Vrba et al. \cite{vrba04}; Nakajima, Tsuji, \&
Yanagisawa \cite{nakajima04}). Hence, T-type dwarfs are all believed
to be brown dwarfs because they have cooled down even below the
minimum atmospheric temperature of the smallest stars.  On the basis
of theoretical predictions and very recent temperature calibrations
(e.g., Dahn et al. \cite{dahn02}), the star--brown dwarf frontier
happens at spectral type L3--L4 at ages of a few Gyr. Field objects
with later types are probably substellar. Various efforts have been
made so far to derive the rotational velocity of L-type dwarfs (e.g.,
Basri et al. \cite{basri00}; Reid et al. \cite{reid02}; Mohanty et al.
\cite{mohanty03}; Bailer-Jones \cite{bailer04}). In contrast to early-
and mid-M stars, L dwarfs appear to be relatively fast rotators with
\vsini~values in the range 10--60\,\kms. To the best of our knowledge,
only the binary T1/T6 $\epsilon$~Indi Bab has a rotation velocity
measurement available in the literature:
\vsini\,=\,28~\kms~(determined for the unresolved system, Smith et al.
\cite{smith03}).

This paper presents high-resolution, near-infrared spectroscopic
observations of 19 bright, ultracool dwarfs with spectral types in the
interval M6.5--T8. Our analysis is intended to first, determine the
rotational velocity of brown dwarfs, with particular emphasis on the
late-L and T-type dwarfs, and second, study the change of rotation
rate from very young ages to the ages typical of the solar
neighborhood, which will allow us to obtain new insight on the angular
momentum evolution in the substellar regime.

\section{The sample, observations and data reduction}

The sample of targets is listed in Table~\ref{obslog}. It comprises 4
late-M type dwarfs, two of which are young and lithium-bearing brown
dwarfs (PPl\,1 is a member of the Pleiades cluster, Stauffer, Schultz,
\& Kirkpatrick \cite{stauffer98}; and LP\,944$-$20 is a likely member
of the Castor moving group, Tinney \cite{tinney98}; Ribas
\cite{ribas03}). The remaining targets are 6 L- and 9 T-type field
dwarfs recently discovered by the 2MASS, SLOAN and Denis surveys;
discovery papers are provided in the last column of
Table~\ref{obslog}. The spectral types given in the second column of
Table are taken from the literature (Kirkpatrick, Henry, \& Simons
\cite{kirk95}; Kirkpatrick, Henry, \& Irwin \cite{kirk97}; Kirkpatrick
et al. \cite{kirk99}, \cite{kirk00}; Mart\'\i n et al.
\cite{martin99}; Geballe et al. \cite{geballe02}; Burgasser et al.
\cite{burgasser06}; Phan-Bao et al.  \cite{phan-bao06}). They were
derived from optical and/or near-infrared low-resolution spectra and
are in the interval M6.5--T8.  There are objects that have been typed
differently by the various groups; we provide all classifications in
Table~\ref{vrotdata}, first the one from Kirkpatrick et al.
\cite{kirk99} or Burgasser et al.  \cite{burgasser06}, second the one
from Geballe et al.  \cite{geballe02}, and finally, the classification
from Mart\'\i n et al. \cite{martin99}. For those targets with the
largest discrepancies (J1632$+$19 and J0255$-$47), we will adopt their
mean spectral type to produce the figures of this paper. In terms of
effective temperature, our sample spans the range 2700--770\,K
(Leggett et al.  \cite{leggett00a}; Vrba et al.  \cite{vrba04}).
J0334$-$49 and vB\,10 can be stars in our sample while the remaining
targets are very likely substellar.

We collected high-resolution near-infrared spectra of the 19 ultracool
dwarfs using the Keck II telescope and the NIRSPEC instrument, a
cross-dispersed, cryogenic echelle spectrometer employing a
1024\,$\times$\,1024 ALADDIN InSb array detector. These observations
were carried out on different occasions from 2000 December through
2006 January and are part of our large program aimed at the study of
radial velocity variability. The log of the observations is shown in
Table~\ref{obslog}. In the echelle mode, we selected the NIRSPEC-3
($J$-band) filter, and an entrance slit width of 0\farcs432 (i.e., 3
pixels along the dispersion direction of the detector), except for
eight targets (J2224$-$01, J1728$+$39, J1632$+$19, J1346$-$00,
J1624$+$00, J1553$+$15, J1217$-$03, and GL\,570D) for which we used an
entrance slit width of 0\farcs576. The length of both slits was
12\arcsec. All observations were performed at an echelle angle of
$\sim$63\arcdeg. This instrumental setup provided a wavelength
coverage from 1.148 up to 1.346\,$\mu$m splitted into ten different
orders, a nominal dispersion ranging from 0.164 (blue wavelengths) to
0.191\,\aa/pix (red wavelengths), and a final resolution element of
0.55--0.70\,\AA~at 1.2485\,$\mu$m (roughly the central wavelength of
the spectra), corresponding to a resolving power
$R$\,$\sim$\,17800--22700. Individual exposure times were function of
the brightness of the targets, ranging from 120 to 900\,s.

\begin{figure}
\epsscale{1.0}
\plotone{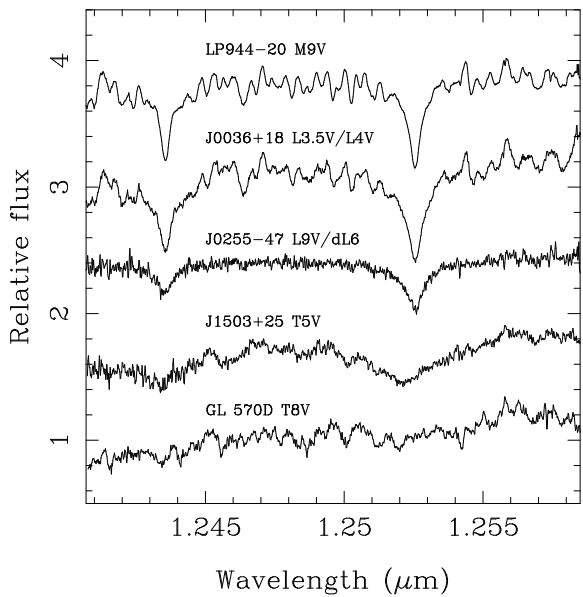}
\caption{NIRSPEC spectra of some of our targets. The most prominent
  features of the M and L dwarfs are due to K\,{\sc i} absorption,
  which vanishes at late-T types. Spectra are normalized to unity in
  the interval 1.2475--1.2490\,$\mu$m and offset by 0.7 on the
  vertical axis. All spectra are shifted in velocity to vacuum
  wavelengths. \label{spec1}}
\end{figure}

Raw data were reduced using the ECHELLE package within
IRAF\footnote{IRAF is distributed by National Optical Astronomy
  Observatory, which is operated by the Association of Universities
  for Research in Astronomy, Inc., under contract with the National
  Science Foundation.}. Spectra were collected at two or three
different positions along the entrance slit. Nodded images were
subtracted to remove sky background and dark current. White-light
spectra obtained with the same instrumental configuration and for each
target were used for flat-fielding the data. All spectra were
calibrated in wavelength using the internal arc lamp lines of Ar, Kr,
and Xe, which were typically acquired after observing the targets. The
vacuum wavelengths of the arc lines were identified and we produced
fits using a third order Legendre polynomial along the dispersion axis
and a second order one perpendicular to it. The mean rms of the fits
was 0.03\,\AA, or 0.7 \kms. We note that any systematic error or
different zero point shifts that may be present in the instrumental
calibration does not affect the rotational velocity measurements. In
order to correct for atmospheric telluric absorptions, near-infrared
featureless stars of spectral types A0--A2 were observed several times
and at different air masses during the various observing runs. The
hydrogen line at 1.282\,$\mu$m, which is intrinsic to these hot stars,
was removed from the spectra before using them for division into the
corresponding science data. Finally, we multiplied the science spectra
by the black body spectrum for the temperature of 9480\,K, which is
adequate for A0V type (Allen \cite{allen00}). We plotted in
Fig.~\ref{spec1} example spectra of our targets around the K\,{\sc i}
doublet at 1.2485\,$\mu$m. Effective temperature (i.e., spectral type)
decreases from top to bottom. Note that all spectra have been shifted
in velocity to vacuum wavelengths for a facile comparison of the
atomic and molecular features. The signal-to-noise ratio of the
spectra varies for different echelle orders and different targets
depending on their brightness. In general, the red orders show better
signal-to-noise ratio than the blue orders, except for the reddest
order centered at 1.337\,$\mu$m in T dwarfs, since these wavelengths
are affected by strong methane and water vapor absorptions below
1300\,K.

\section{Rotational velocities}

\begin{figure}
\epsscale{1.0}
\plotone{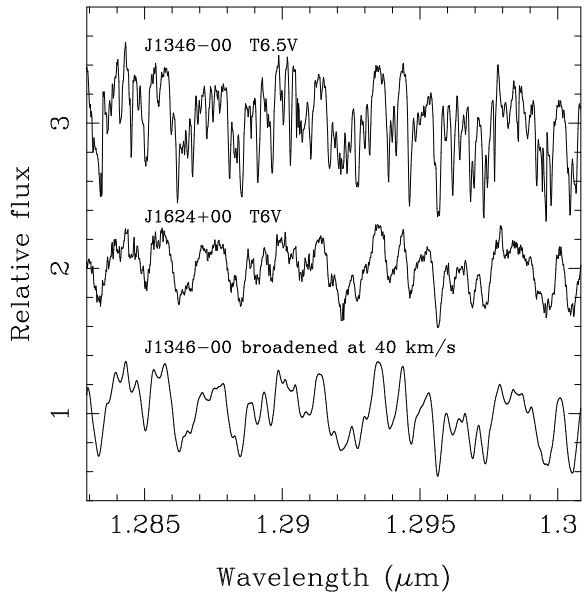}
\caption{NIRSPEC spectra of the brown dwarf J1346$-$00 (T6.5V, low \vsini, top)
  and the rapid rotator J1624$+$00 (T6V, middle). The spectrum of J1624$+$00
  has been shifted in velocity to match J1346$-$00 for a facile comparison of
  features. The spectrum of J1346$-$00 convolved with a rotation profile of
  40~\kms~is shown at the bottom. Data are scaled to a common vertical range
  and offset by unity on the vertical axis for clarity. \label{spec2}}
\end{figure}

We measured projected rotational velocities via a cross-correlation of
the spectra of our targets against spectra of slowly rotating dwarfs
of similar types. The details of the procedure, which is based on the
assumption that the line broadening is dominated by rotation, is fully
described in the literature (e.g., Reid et al. \cite{reid02};
Bailer-Jones \cite{bailer04}, and references therein). Summarizing,
rotational velocities, \vsini, are obtained from the width of the peak
of the cross-correlation function between the targets and the slow
rotator templates. The larger the width, the faster the rotation rate.
We performed the cross-correlation on each order separately, fit a
Gaussian to the peak, and measured its width, which is calibrated
against \vsini~by applying artificial broadening for a range of
velocities (10 up to 50 \kms~in steps of 5 \kms) to the spectra of
slow rotators. We adopted the line rotation profile given by Gray
\cite{gray92}, and a limb darkening parameter of $\epsilon$\,=\,0.6.
One example of the artificially broadened spectra is displayed in
Fig.~\ref{spec2} (bottom spectrum). Only orders for which we
unambiguously identified the peak of the cross-correlation function
and obtained good fits were used. Then, all results were averaged to
produce the final \vsini~measurement. Uncertainties are derived from
the standard deviation of the mean.

Ideally, non-rotating templates should be used. Mohanty \& Basri
\cite{mohanty03} reported a very low projected velocity of the
M8V-type star vB\,10 (\vsini\,=\,6.5 \kms). This is well below the
instrumental profile broadening of our data, which is about
15\,\kms~based on one pixel resolution, and is also smaller than the
minimum detectable rotation from our data (see below). As a result,
vB\,10 can be considered a non-rotator in our study, and has been
employed as a template to obtain rotational velocities of the M- and
L-type targets.  The cross-correlation technique assumes that the
target and template spectra are of similar type and differ only in the
rotation velocity.  Nevertheless, Bailer-Jones \cite{bailer04} has
recently shown that M-type templates can also yield accurate rotation
velocities of L dwarfs. In our sample, the energy distribution of T
dwarfs do differ significantly from M dwarfs, and we found that for
the T-type targets the cross-correlation with vB\,10 gives reasonable
results if resonance lines of K\,{\sc i} are avoided and only
molecular (particularly water vapor) lines are used in the
cross-correlation.  The cool brown dwarf J1346$-$00 (T6.5V) turns out
to have a low value of \vsini~(because of the unknown inclination
angle of the rotation axis, we cannot establish whether J1346$-$00 is
a true fast or slow rotator. However, we used it as a ``slow rotator''
in our study). From the data we can impose an upper limit of 15
\kms~on its projected rotational velocity. We used J1346$-$00 as a
secondary template to measure rotational broadening of the coolest
objects in our sample, the late-L and T-type dwarfs. We note, however,
that using a rotating template will underestimate the rotation
velocities (Mohanty \& Basri \cite{mohanty03} provide an extensive and
quantitative discussion on this respect). Our derived
\vsini~measurements using vB\,10 and J1346$-$00 as the slowly rotating
reference objects are given in columns~3 and~4 of
Table~\ref{vrotdata}, respectively. The agreement between the two
columns is good within 1-$\sigma$ the uncertainty (i.e., there is no
obvious tendency for \vsini~obtained using J1346$-$00 to be lower than
the values derived with the M-type dwarf).  For those objects in
common with other studies (2 M dwarfs and 4 L dwarfs), we also give
their previously derived projected rotational velocities in column~5
of Table~\ref{vrotdata}. With the only exception of J0036$+$18 (see
below), our measurements and the values from the literature agree
within the expected errors.

Figure~\ref{spec2} depicts the spectra of the T-type secondary
template (J1346$-$00) and of one rapid rotator of similar
classification (J1624$+$00, T6V). The data of J1624$+$00 were
shifted in velocity to match the template. For comparison purposes,
the spectrum of J1346$-$00 broadened at the (roughly) rotation
velocity of J1624$+$00 is also shown at the bottom of the Figure. The
resemblance of J1624$+$00 to the artificially broadened spectrum is
remarkable, supporting the \vsini~results of the cross-correlation
technique.

In our sample of 19 ultracool dwarfs, only 3 of them have small values
of \vsini: the reference M8-type dwarf vB\,10, the nearby M9-dwarf
J0334$-$49, and the T6.5-type object J1346$-$00. For the latter two we
have placed an upper limit of 15 \kms. The cross-correlation technique
suggests that the M9-dwarf rotates at a speed of 8.5\,$\pm$\,2.6 \kms.
However, this measurement is below the minimum detectable projected
rotational velocity of our study, which we estimate at about
10.5~\kms~after the discussion by Bailer-Jones \cite{bailer04}.
Therefore, the adoption of the mean NIRSPEC instrumental broadening as
the upper limit on the \vsini~for the slowest rotators in our sample
is conservative. It may be noteworthy that we have found a slowly
rotating T dwarf in a sample of 9 T-type dwarfs, while a much larger
sample of L dwarfs drawn from our work and the literature has yet to
yield one. However, we note that there are a few L-dwarfs with
\vsini~$\sim$~10 \kms, and that the rotation inclinations are randomly
distributed.

\section{Discussion}

\subsection{Periodic photometric variability and \vsini}

Four L-type dwarfs among our targets have tentative rotational periods
available in the literature. Berger et al. \cite{berger05} reported on
the detection of strongly variable and periodic radio emission in
J0036$+$18 (L3.5V) with a periodicity of 3\,h. These authors suggested
various scenarios to account for the observed periodic radio emission,
one of which is related to equatorial rotation and an anchored,
long-lived magnetic field. The period of 3\,h is consistent with our
\vsini~measurement (36.0 \,$\pm$\,2.7 \kms); thus, it is plausibly the
true rotation periodicity. Using a radius of 0.09~$R_{\odot}$, the
inferred inclination angle of the rotation axis is
$i$\,$\ge$\,66\arcdeg~with an expectation value of 81\arcdeg. We note,
however, that our \vsini~differs by a factor of 2.4 with respect the
measurement given by Schweitzer et al. \cite{schweitzer01}. These
authors employed a different technique to determine rotation
velocities: they compared artificially broadened theoretical spectra
to observed spectra. Using a similar approach, Jones et al.
\cite{jones05} obtained a rotation velocity (see Table~\ref{vrotdata})
in better agreement with our measurement.

Recently, Koen \cite{koen05} has discussed the possible variable nature of
J0255$-$47 (L9V/dL6) on more than one timescale. This author found two
possible photometric modulations with periodicities of 1.7 and 5.2 h in the
optical light curve of J0255$-$47. If this object has a radius of
0.09~$R_{\odot}$, the longest period may be ruled out as the rotation
period. However, the shortest periodicity is consistent with a rotational
modulation, and when combined with our \vsini~would yield a rotation axis
inclination angle of $i$\,$\sim$\,40\arcdeg. Then, the true equatorial
velocity of J0255$-$47 would turn out to be $\sim$64 \kms. 

The L5V dwarf J0539$-$00 has also been detected to be photometrically
variable by Bailer-Jones \& Mundt \cite{bailer01}, with a derived
periodicity of 13.3 h. And a tentative modulation of 21.8 h associated
to the L4.5V dwarf J2224$-$01 has been reported by Gelino et al.
\cite{gelino02}. Given our \vsini~data and the radius of
0.09~$R_{\odot}$, such long periods cannot be securely ascribed to
rotation modulation of a long-lived atmospheric feature. Similarly,
other ultracool dwarfs show too long periodic variability to be
related to the rotation period (e.g., Mart\'\i n, Zapatero Osorio, \&
Lehto \cite{martin01}; Gelino et al.  \cite{gelino02}). Nevertheless,
long periodicities could be interpreted as due to equatorial rotation
if these objects had larger size, i.e., were significantly younger
than the rest of the objects in the solar neighborhood.

\subsection{Rotation and angular momentum evolution}

From Table~\ref{vrotdata}, field T-type brown dwarfs are relatively
fast rotators with \vsini~measurements in the interval $\le$15--40
\kms. All state-of-the-art evolutionary models available in the
literature predict that the size of brown dwarfs keeps approximately
constant (0.08--0.1~$R_{\odot}$) at ages older than $\sim$0.5\,Gyr
regardless the mass. Under the assumption that this theoretical
prediction is correct, the rotation periods of T-type brown dwarfs are
probably below 12.5\,h. To date, only a few T dwarfs have been
photometrically monitored to search for rotational modulation (e.g.,
Burgasser et al. \cite{burgasser02a}; Goldman \cite{goldman03}; Koen
\cite{koen05}). Enoch, Brown, \& Burgasser \cite{enoch03} found
evidence for periodic behavior on a timescale of 3\,h in one T1 brown
dwarf. Unfortunately, this variable object is not in our sample;
nevertheless, the derived photometric periodicity is consistent with
our prediction for T-type brown dwarfs based on the radii given by the
evolutionary models.

\begin{figure}
\epsscale{1.0}
\plotone{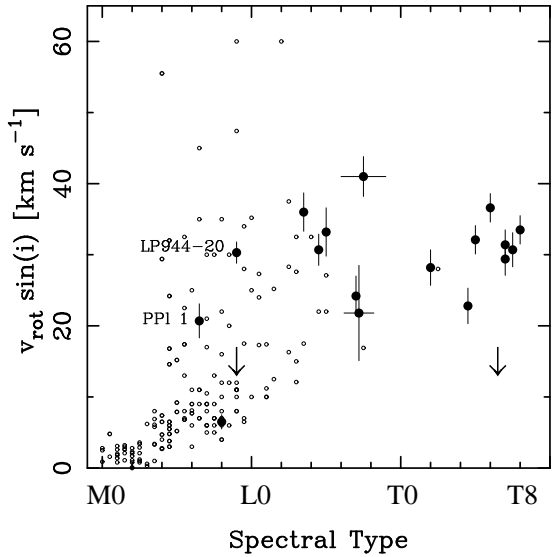}
\caption{Spectroscopic rotation velocities of M-, L- and T-type
  dwarfs. Our sample is plotted as filled circles and arrows denoting
  positive detections and upper limits, respectively. The two known
  young dwarfs of our sample are labeled. The datum of vB\,10 (M8V),
  one of the template sources for rotation velocity measurement, is
  taken from Mohanty \& Basri \cite{mohanty03}. Two L dwarfs show
  large error bars in spectral type to account for their different
  classifications found in the literature. Small open circles
  correspond to data (positive detections) from the literature (see
  text). We note that a large fraction of M0--M4 stars have
  non-detections (\vsini $\le$ 5 \kms), and their upper limits are not
  plotted here for clarity.  \label{vrot}}
\end{figure}

Our \vsini~results are plotted against spectral type in
Fig.~\ref{vrot}, where we have also included data of field M stars and
L dwarfs from the literature (Marcy \& Chen \cite{marcy92}; Basri \&
Marcy \cite{basri95}; Delfosse et al. \cite{delfosse98}; Basri et al.
\cite{basri00}; Reid et al. \cite{reid02}; Mohanty \& Basri
\cite{mohanty03}; Bailer-Jones \cite{bailer04}). The
\vsini~measurement of the unresolved T-dwarf binary $\epsilon$~Indi
Bab obtained by Smith et al. \cite{smith03} is also plotted in the
Figure. Note that many early-M to mid-M stars have very slow rotation
(\vsini\,$\le$\,5 \kms, Delfosse et al. \cite{delfosse98}; Mohanty \&
Basri \cite{mohanty03}), and for the sake of clarity we have not
plotted in the Figure the upper limits on their detection. It is
obvious that there is a correlation between rotation and spectral
type: rotational velocity keeps increasing with later spectral type
from M to T types. This behavior was previously reported for M and L
dwarfs in the literature (Mohanty \& Basri \cite{mohanty03};
Bailer-Jones \cite{bailer04}), and is in part due to the different
mass of the objects contemplated in Fig.~\ref{vrot}. We will not
consider the young brown dwarfs of our sample, PPl\,1 and
LP\,944$-$20, in the following discussion. At the typical age interval
of the solar vicinity ($\sim$1--10\,Gyr), T-type objects are
substellar with likely masses in the brown dwarf regime (20--70 \Mj,
Baraffe et al.  \cite{baraffe03}). On the contrary, and for the same
age interval, low-mass and very low-mass field stars above the
substellar limit have M and early-L types. From Fig.~\ref{vrot}, and
despite the uncertainty introduced by the inclination angle of the
rotation axis, field brown dwarfs indeed rotate more rapidly than
field M-type stars. Scholz \& Eisl\"offel \cite{scholz04} and
\cite{scholz05} studied the rotation period--mass relationship of
stellar and substellar members in the $\sigma$~Orionis and
$\epsilon$~Orionis young clusters finding that the lower the mass, the
shorter the rotation period.

Field L-type dwarfs can be either very low-mass stars with probable
masses below 0.1~$M_{\odot}$ or brown dwarfs depending on age and
surface temperature. About 35\%~of L3V--L4V dwarfs have lithium in
their spectra, and the rate increases up to 70\%~at L6V (Kirkpatrick
et al. \cite{kirk00}). These are definitively substellar with likely
masses below 60~\Mj. L-type objects are also reported in the
literature to be fast rotators (Basri et al. \cite{basri00};
Bailer-Jones \cite{bailer04}), with projected rotational velocities up
to 60 \kms, and a minimum velocity of 10 \kms~(Mohanty \& Basri
\cite{mohanty03}). From our data, T-type brown dwarfs do not appear to
have significantly faster rotation than L dwarfs, and a smooth
transition between the L- and T-types can be seen in Fig.~\ref{vrot}.
However, we note that there is a hint for the lower envelope of the
\vsini--spectral type relation to increase linearly from mid-M to T
types. This trend may be an artifact of the small sample size of T
dwarfs and of our detection limit; hence, more measurements of
different T-type brown dwarfs would be required to definitively
ascertain this tendency.

Also relevant is the exploration of the angular momentum evolution of
brown dwarfs. To address this issue we should know the age of our
sample. PPl\,1 is a Pleiades brown dwarf (60--72~\Mj) with an
estimated age of 120\,Myr. We have calculated the space motions of all
the other targets except for two T dwarfs that lack parallax
measurements (forthcoming paper), and used the criteria defined by
Leggett \cite{leggett92} to group them into young disk, young-old disk
and old disk kinematic categories. LP\,944$-$20 was widely discussed
by Ribas \cite{ribas03}. According to their galactic velocities, all L
and T dwarfs of Table~\ref{vrotdata} can be classified as members of
the young and young-old disk, for which we have assumed the following
age intervals: 0.6--1.5\,Gyr (young disk) and 1--5\,Gyr (young-old
disk). We note that Reid et al. \cite{reid02} also pointed out that
late-M dwarfs appear kinematically younger than early-M stars, and
that young M dwarfs rotate faster than the older ones. From our data,
we find no difference in \vsini~between the young disk T dwarfs and
the young-old disk ones.

\begin{figure}
\epsscale{1.0}
\plotone{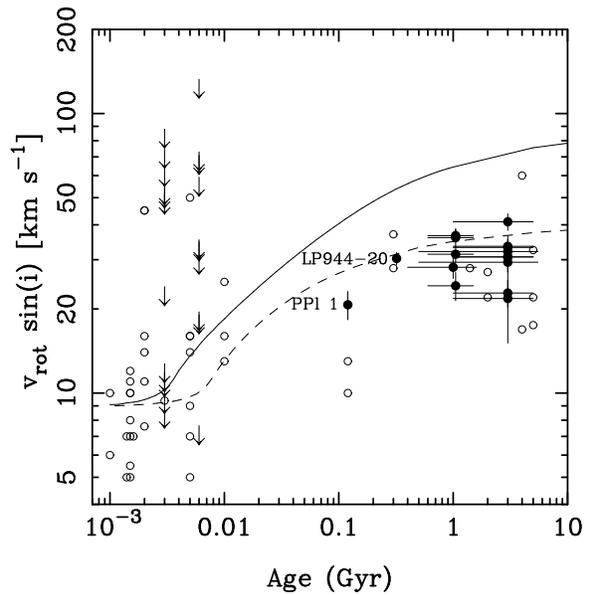}
\caption{Projected rotation velocity as a function of age for brown
  dwarfs with likely masses in the interval 30--70~\Mj. Our data
  (PPl\,1, LP\,944$-$20, and field dwarfs later than L3.5V) are
  plotted as filled circles. Data from the literature (see text for
  details) are shown with open circles and arrows (upper limits). The
  curves trace the hydrostatic contraction of brown dwarfs with 
  60~\Mj~(solid line) and 30~\Mj~(dashed line), which have a rotation 
  velocity of 9~\kms~at the age of 1\,Myr.
  \label{vrot_age}}
\end{figure}

Our \vsini~values are plotted against age in Fig.~\ref{vrot_age},
where we have considered only brown dwarfs (PPl\,1, LP\,944$-$20, and
the field dwarfs cooler than L3.5V). Further spectroscopic rotation
velocities of confirmed lithium-bearing Pleiades brown dwarfs and of
the substellar pair GJ\,569Ba and GJ\,569Bb are obtained from the
literature (Basri \& Marcy \cite{basri95}; Mart\'\i n et al.
\cite{martin98}; Zapatero Osorio et al. \cite{osorio04}), and are also
shown in the Figure. Despite the scarce number of data for ages of
hundred Myr, there is no clear evidence for a significant braking
between the Pleiades age and the field. This contrasts markedly to
solar-type stars. At the age of the Hyades ($\sim$600 Myr), most G-
and K-type stars are rather slow rotators, and some M-type stars
display relatively rapid rotation with maximum
\vsini\,=\,20\,\kms~(Stauffer et al. \cite{stauffer97}).

We have compiled the \vsini~measurements of very young brown dwarfs
with ages between $\sim$1 and 10~Myr from Joergens \& Guenther
\cite{joergens01}, Muzerolle et al. \cite{muzerolle03}, and Mohanty,
Jayawardhana \& Basri \cite{mohanty05}. These data correspond to
confirmed substellar members of the following star-forming regions and
young clusters: $\rho$ Ophiuchi ($\sim$1~Myr), Taurus ($\sim$1.5~Myr),
IC\,348 ($\sim$2~Myr), Chamaeleon~I ($\sim$2~Myr), $\sigma$~Orionis
($\sim$3~Myr), Upper Scorpius ($\sim$5~Myr), and TW Hya
($\sim$10~Myr). It is accepted that the substellar limit is located at
about M6--M6.5 spectral type at these young ages as well as in the
Pleiades (e.g., Luhman et al. \cite{luhman98}; Mart\'\i n et al.
\cite{martin98}). Hence, to secure that only brown dwarfs are depicted
in Fig.~\ref{vrot_age}, we have plotted sources with spectral type M7
and later (the coolest brown dwarf of these spectroscopic studies has
spectral class M9.5).  Their masses are likely in the interval
30--70~\Mj, i.e., fully overlapping with the mass estimates for the
field T-type objects of our sample. The projected rotation velocities
of the very young ($\le$10~Myr) brown dwarfs span the range
0--60~\kms.  In addition, several groups have photometrically
monitored substellar members of the $\sigma$~Orionis and
$\epsilon$~Orionis ($\sim$6~Myr) clusters (Bailer-Jones \& Mundt
\cite{bailer01}; Scholz \& Eisl\"offel \cite{scholz04},
\cite{scholz05}; Zapatero Osorio et al.  \cite{osorio03}; Caballero et
al. \cite{caballero04}), finding periodic behaviors on timescales
between 3 and 80\,h. The various groups of authors discuss that the
observed modulated variability is possibly related to rotation in many
cases. We have used the radii predictions of Baraffe et al.
\cite{baraffe03} to transform periods into velocities, which are
plotted as arrows (denoting upper limits) in Fig.~\ref{vrot_age}. From
the Figure, it is noteworthy that a significant amount ($\sim$50\%) of
young, massive brown dwarfs have \vsini\,$\le$\,10~\kms, while, in
general, field substellar objects of similar mass rotate three to six
times faster. Brown dwarfs definitively undergo a spin-up because of
gravitational contraction, and their spin-down time appears to be
longer than that of solar-type stars.  Chabrier \& K\"uker
\cite{chabrier06} argued that the magnetic field topology of fully
convective objects exhibits an asymmetric structure, which in addition
to the decreasing conductivity in the atmosphere of ultracool dwarfs
reduces the efficiency of magnetic braking, providing a possible
explanation for the lower angular momentum loss of brown dwarfs.

Nevertheless, fast-rotating brown dwarfs do exist at all ages from a few Myr
through several Gyr. This is independent of evolutionary models, since
rotation periods of a few hours and \vsini\,$\ge$\,10~\kms~are equally
observed among young and field brown dwarfs. This suggests some angular
momentum loss throughout the lifetime of a brown dwarf. Furthermore, with the
current evolutionary models, simple hydrostatic contraction (i.e., angular
momentum conservation) fails to reconcile the entire \vsini~distributions of
the young ($\le$10~Myr) and ``old'' ($\ge$100~Myr) substellar populations
shown in Fig.~\ref{vrot_age}. As an example, we have plotted in the
\vsini--age diagram of Fig.~\ref{vrot_age} the purely hydrostatic contraction
of two presumed brown dwarfs with masses of 30 (dashed line) and 60~\Mj~(solid
line) by assuming a moderate ``initial'' rotation velocity of 9\,\kms~and the
theoretical models of Baraffe et al. \cite{baraffe03}. We note that 60~\Mj~is
probably the most representative mass of the objects in
Fig.~\ref{vrot_age}. At the very young ages, there are brown dwarfs located
above and below the curves of angular momentum conservation; however, for ages
older than 1~Gyr, nearly all brown dwarfs lie below the curves. Hence, some
braking becomes apparent. We remark that the conclusions drawn from this
discussion are based on current evolutionary models, and should be revised if
the models are proved to be wrong.

The fastest rotating young brown dwarfs with masses in the interval
50--72~\Mj, periods below 10\,h, and $v_{\rm
  rot}$\,$\ge$\,45~\kms~(note the drop of sin\,$i$) pose a problem to
theory (Scholz et al. \cite{scholz05b}). Relying on actual models,
these rapid, massive substellar rotators would reach the spherical
breakup velocity in less than the lifetime of the Galaxy unless they
brake down dramatically. It is expected that fast rotation impacts the
evolution of brown dwarfs. These objects have rather neutral
atmospheres (low conductivity) and despite the fact they can hold some
long-lived magnetic fields up to a few Gyr (Berger et al.
\cite{berger05}, and references therein), the atmospheric fluid
motions and the field are likely decoupled, which reduces the
``magnetic'' activity. The mechanisms through which ultracool
(particularly the L and T types), fully convective rotators can lose
angular momentum are to be determined (magnetic braking, frictional
heating, etc.). However, it is worth mentioning that young massive
brown dwarfs of mid- to late-M spectral classes are warm enough to
have coronae in a similar manner than low-mass stars (see, e.g.,
Preibisch \& Zinnecker \cite{preibisch02}; Mokler \& Stelzer
\cite{mokler02}; Bouy \cite{bouy04}; Preibisch et al.
\cite{preibisch05}, and references therein). This could in principle
provide a braking mechanism but at a time when the brown dwarfs have
not yet fully contracted and have not cooled down significantly.
Because there is no obvious braking between the Pleiades and field
brown dwarfs (Fig.~\ref{vrot_age}), it might be possible that these
objects spin up due to contraction and brake down while still being
warm (and very young), and once they evolve to cooler temperatures the
braking time becomes longer. Very recently, Reiners \& Basri
\cite{reiners06} report on the high \vsini~of an halo L-dwarf,
suggesting that braking is even inexistent (provided the object has an
``halo'' age).  Additionally, flare activity, like the one reported in
a few field L dwarfs (e.g., Liebert et al.  \cite{liebert03}) can also
shed off angular momentum.  Nevertheless, the frequency of strong
flares in cool objects is quite low ($\le$7\%, Gizis et al.
\cite{gizis00}), so their contribution to the rotation braking is
expected to be small. Finally, we point out that there can be a
diversity of angular momentum histories in our sample of brown dwarfs;
some could be binaries, some may have flares, etc.

\section{Summary and conclusions}

We obtained high-resolution ($R$\,$\sim$\,17800--22700), near-infrared
(1.148--1.346\,$\mu$m) spectra of a sample of 19 ultracool dwarfs
(M6.5--T8) using the NIRSPEC instrument and the 10-m Keck II
telescope. Our sample comprises two late-M dwarfs, two young
(100--500\,Myr) brown dwarfs and fifteen L3.5--T8 likely brown dwarfs
of the solar neighborhood. Projected rotation velocities (\vsini) were
measured via a cross-correlation of the data against spectra of
slow-rotating dwarf templates of similar types, which were observed
with the same instrumental configuration. The star vB\,10 (M8V) and
the brown dwarf J1346$-$00 (T6.5V) were the templates. We found that,
in general, the T-type brown dwarfs are relatively fast rotators, as
are also the L dwarfs, with \vsini~spanning the range
$\le$15--40~\kms. Assuming a radius of 0.08--0.1\,$R_{\odot}$, we
placed an upper limit of 12.5\,h on the equatorial rotation of T
dwarfs.  Rotation appears to correlate with spectral type (from M to T
class), \vsini~increasing towards later types. However, we found no
clear evidence for T dwarfs rotating significantly faster than L
dwarfs.

We compared our \vsini~measurements to photometric periods of four L
dwarfs reported in the literature, and concluded that the 3-h periodic
radio emission of J0036$+$18 (Berger et al. \cite{berger05}) is likely
the true rotation period of the object. However, the derived
photometric periodicities of J0255$-$47 (5.2\,h, Koen \cite{koen05}),
J0539$-$00 (13.3\,h, Bailer-Jones \& Mundt \cite{bailer01}), and
J2224$-$01 (21.8\,h, Gelino et al. \cite{gelino02}) seem too long to
be considered the rotation periods if we assume a radius of
0.09\,$R_{\odot}$.

We also compared our data to \vsini~measurements of brown dwarfs of
similar mass (30--72~\Mj), which are confirmed members of very young
star-forming regions and star clusters (1--10\,Myr). A significant
fraction of the very young brown dwarfs show \vsini~$\le$~10 \kms,
which contrasts with the high values found in field brown dwarfs. This
supports the fact that brown dwarfs spin up with age due to
gravitational contraction. From the calculation of the hydrostatic
contraction based on radii predictions of state-of-the-art
evolutionary models, we conclude that brown dwarfs seem to brake down
from 1\,Myr to the age of the solar neighborhood, and hence, they lose
some angular momentum.  However, massive brown dwarfs do not appear to
brake down as dramatically as solar-type to early-M stars, suggesting
that their spin-down time is significantly longer.

\acknowledgements We thank V. J. S. B\'ejar for useful discussions. We
also thank G. Basri (the referee of this paper) for his comments and
suggestions.  This research has been supported by a Keck PI Data
Analysis grant awarded to E. L. M. by Michelson Science Center, and by
the Spanish projects AYA2003-05355 and AYA2005-06453.  We thank the
Keck observing assistants and the staff in Waimea for their kind
support.  The data were obtained at the W. M. Keck Observatory, which
is operated as a scientific partnership between the California
Institute of Technology, the University of California, and NASA. The
Observatory was made possible by the generous financial support of the
W. M. Keck Foundation. The authors extend special thanks to those of
Hawaiian ancestry on whose sacred mountain we are privileged to be
guests. This research has made use of the SIMBAD database, operated at
CDS, Strasbourg, France.

\clearpage
\begin{deluxetable}{lcccccc}
\tabletypesize{\scriptsize}
\tablecaption{Keck/NIRSPEC observing log. \label{obslog}}
\tablewidth{0pt}
\tablehead{
\colhead{Object}             & \colhead{Obs$.$ date} & \colhead{Slit name} & \colhead{Echelle} & \colhead{Exp$.$ time} & \colhead{Airmass} & Reference\\
\colhead{      }             & \colhead{           } & \colhead{         } & \colhead{(deg)  } & \colhead{(s)        } & \colhead{       } & \colhead{}
}
\startdata
PPl\,1                       & 2005 Oct 26 & 0.432$\times$12 & 63.00   & 3$\times$480 & 1.01    & 1  \\
                             & 2005 Oct 28 & 0.432$\times$12 & 63.00   & 3$\times$420 & 1.02    &    \\
vB10                         & 2001 Jun 15 & 0.576$\times$12 & 63.00   & 2$\times$300 & 1.14    & 2  \\
                             & 2001 Nov 02 & 0.432$\times$12 & 63.29   & 8$\times$120 & 1.40    &    \\
DENIS-P\,J033411.39$-$495333.6\tablenotemark{a} & 2005 Oct 26 & 0.432$\times$12 & 63.00   & 6$\times$300 & 2.86 &  3  \\
                             & 2005 Oct 28 & 0.432$\times$12 & 63.00   & 3$\times$300 & 2.86    &    \\
LP\,944$-$20                 & 2000 Dec 12 & 0.432$\times$12 & 63.00   & 3$\times$200 & 1.89    & 4  \\
                             & 2005 Oct 26 & 0.432$\times$12 & 63.00   & 6$\times$300 & 1.86    &    \\
                             & 2005 Oct 27 & 0.432$\times$12 & 63.00   & 6$\times$300 & 1.75    &    \\
                             & 2005 Oct 28 & 0.432$\times$12 & 63.00   & 6$\times$300 & 1.76    &    \\
                             & 2005 Oct 29\tablenotemark{b} & 0.432$\times$12 & 63.00   & 1$\times$300 & 1.80 &    \\
2MASS\,J00361617$+$1821104   & 2004 Dec 05 & 0.432$\times$12 & 63.00   & 6$\times$300 & 1.06    & 5  \\
                             & 2005 Oct 28 & 0.432$\times$12 & 63.00   & 3$\times$300 & 1.11    &    \\
2MASS\,J22244381$-$0158521   & 2001 Jun 15 & 0.576$\times$12 & 63.00   & 2$\times$600 & 1.09    & 6  \\
SDSS\,J053951.99$-$005902.0  & 2001 Nov 02 & 0.432$\times$12 & 62.78   & 8$\times$300 & 1.30    & 7  \\
                             & 2005 Oct 27 & 0.432$\times$12 & 63.00   & 3$\times$480 & 1.15    &    \\
2MASS\,J17281150$+$3948593AB & 2001 Jun 15 & 0.576$\times$12 & 63.00   & 2$\times$900 & 1.33    & 6  \\
2MASS\,J16322911$+$1904407   & 2001 Jun 15 & 0.576$\times$12 & 63.00   & 2$\times$900 & 1.37    & 8  \\
DENIS-P\,J0255.0$-$4700      & 2005 Oct 27 & 0.432$\times$12 & 63.00   & 3$\times$420 & 2.53    & 9  \\
SDSSp\,J125453.90$-$012247.4 & 2006 Jan 19 & 0.432$\times$12 & 63.00   & 9$\times$600 & 1.09    & 10 \\
2MASS\,J05591914$-$1404488   & 2001 Nov 02 & 0.432$\times$12 & 62.78   & 8$\times$300 & 1.39    & 11 \\
                             & 2004 Dec 05 & 0.432$\times$12 & 63.00   & 2$\times$300 & 1.36    &    \\
                             & 2005 Oct 26 & 0.432$\times$12 & 63.00   & 3$\times$480 & 1.18    &    \\
                             & 2005 Oct 27 & 0.432$\times$12 & 63.00   & 3$\times$480 & 1.27    &    \\
                             & 2005 Oct 28 & 0.432$\times$12 & 63.00   & 3$\times$480 & 1.24    &    \\
2MASS\,J15031961$+$2525196   & 2006 Jan 19 & 0.432$\times$12 & 63.00   & 2$\times$300 & 1.07    & 12 \\
SDSS\,J162414.37$+$002915.6  & 2001 Jun 15 & 0.576$\times$12 & 63.00   & 2$\times$900 & 1.13    & 13 \\
SDSS\,J134646.45$-$003150.4  & 2001 Jun 15 & 0.576$\times$12 & 63.00   & 3$\times$900 & 1.14    & 14 \\
2MASS\,J15530228$+$1532369   & 2001 Jun 15 & 0.576$\times$12 & 63.00   & 2$\times$900 & 1.04    & 15 \\
2MASS\,J12171110$-$0311131   & 2001 Jun 15 & 0.576$\times$12 & 63.00   & 2$\times$900 & 1.15    & 16 \\
GL\,570D                     & 2001 Jun 15 & 0.576$\times$12 & 63.00   & 2$\times$900 & 1.45    & 17 \\
2MASS\,J04151954$-$0935066   & 2005 Oct 26 & 0.432$\times$12 & 63.00   & 2$\times$600 & 1.18    & 15 \\
\enddata
\tablenotetext{a}{It is also LEHPM\,3396 (Pokorny, Jones, \& Hambly \cite{pokorny03}).}
\tablenotetext{b}{Cloudy, low transparency. Not used for \vsini~determination.}
\tablerefs{(1) Stauffer et al. \cite{stauffer98}; (2) Van Biesbroeck \cite{van61}; (3) Pokorny et al. \cite{pokorny03}; (4) Tinney \cite{tinney98}; (5) Reid et al. \cite{reid00}; (6) Kirkpatrick et al. \cite{kirk00}; (7) Fan et al. \cite{fan00}; (8) Kirkpatrick et al. \cite{kirk99}; (9) Mart\'\i n et al. \cite{martin99}; (10) Leggett et al. \cite{leggett00b}; (11) Burgasser et al. \cite{burgasser00a}; (12) Burgasser et al. \cite{burgasser03}; (13) Strauss et al. \cite{strauss99}; (14) Tsvetanov et al. \cite{tsvetanov00}; (15) Burgasser et al. \cite{burgasser02b}; (16) Burgasser et al. \cite{burgasser99}; (17) Burgasser et al. \cite{burgasser00b}.}
\end{deluxetable}

\clearpage

\begin{deluxetable}{lccccc}
\tabletypesize{\scriptsize}
\tablecaption{Spectroscopic rotation velocities of our sample. \label{vrotdata}}
\tablewidth{0pt}
\tablehead{
\multicolumn{2}{c}{} & \multicolumn{2}{c}{Template object} \\
\cline{3-4} \\
\colhead{Object}               & \colhead{SpT} & \colhead{vB\,10       } & \colhead{J1346$-$00}    & Previous & Reference\\
\colhead{      }               & \colhead{   } & \colhead{(km s$^{-1}$)} & \colhead{(km s$^{-1}$)} & \colhead{(km s$^{-1}$)} & \\
}
\startdata
PPl\,1                         & M6.5               & 20.7$\pm$2.4 & \nodata      & 18.5$\pm$1.5    & 1     \\
vB\,10                         & M8V                & \nodata      & \nodata      & 6.5             & 2     \\
DENIS-P\,J033411.39$-$495333.6 & M9                 & $\le$15      & \nodata      &                 &       \\
LP\,944$-$20                   & M9V                & 30.3$\pm$1.5 & \nodata      & 30.0$\pm$2.5, 28, 28.3, 31 & 2,3,4,5,6 \\
2MASS\,J00361617$+$1821104     & L3.5V/L4V          & 36.0$\pm$2.7 & \nodata      & 38, 15          & 6,7   \\
2MASS\,J22244381$-$0158521     & L4.5V              & 30.3$\pm$3.4 & 31.1$\pm$2.2 & 20.9--29.2      & 8     \\
SDSS\,J053951.99$-$005902.0    & L5V                & 32.3$\pm$3.6 & 34.0$\pm$3.4 &                 &       \\
2MASS\,J17281150$+$3948593AB   & L7V                & 25.1$\pm$7.2 & 23.3$\pm$2.8 &                 &       \\
2MASS\,J16322911$+$1904407     & L8V/L7.5V/dL6      & 22.6$\pm$6.7 & 20.9$\pm$7.0 & 30$\pm$10       & 2,3   \\
DENIS-P\,J0255.0$-$4700        & L9V/ /dL6          & 40.8$\pm$8.0 & 41.1$\pm$2.8 & 40$\pm$10       & 2,3   \\
SDSSp\,J125453.90$-$012247.4   & T2V                & 29.5$\pm$5.0 & 27.3$\pm$2.5 &                 &       \\
2MASS\,J05591914$-$1404488     & T4.5V              & 25.4$\pm$5.0 & 20.1$\pm$4.8 &                 &       \\
2MASS\,J15031961$+$2525196     & T5V                & 31.4$\pm$5.0 & 32.8$\pm$2.0 &                 &       \\
SDSS\,J162414.37$+$002915.6    & T6V                & 34.6$\pm$5.0 & 38.5$\pm$2.0 &                 &       \\
SDSS\,J134646.45$-$003150.4    & T6.5V              & $\le$15      & \nodata      &                 &       \\
2MASS\,J15530228$+$1532369AB   & T7V                & \nodata      & 29.4$\pm$2.3 &                 &       \\
2MASS\,J12171110$-$0311131     & T7V/T8V            & \nodata      & 31.4$\pm$2.1 &                 &       \\
GL\,570D                       & T7.5V/T8V          & 32.8$\pm$5.0 & 28.6$\pm$2.4 &                 &       \\
2MASS\,J04151954$-$0935066     & T8V                & \nodata      & 33.5$\pm$2.0 &                 &       \\
\enddata
\tablecomments{Spectroscopic rotation velocities are
  \vsini~measurements.}  
\tablerefs{(1) Mart\'\i n et al.
  \cite{martin98}; (2) Mohanty \& Basri \cite{mohanty03}; (3) Basri et
  al.~\cite{basri00}; (4) Reid et al. \cite{reid02}; (5) Tinney \&
  Reid \cite{tinney98b}; (6) Jones et al. \cite{jones05}; (7)
  Schweitzer et al. \cite{schweitzer01}; (8) Bailer-Jones
  \cite{bailer04}.}
\end{deluxetable}

\clearpage

\end{document}